\documentclass[a4paper,11pt]{article}
\pdfoutput=1 
\usepackage{jcappub}

\usepackage{graphicx}
\usepackage{bm}
\usepackage{hyperref}
\usepackage{color}

\definecolor{orange}{rgb}{1,0.5,0}

\def\be{\begin{equation}}
\def\ee{\end{equation}}

\def\gtsima{$\; \buildrel > \over \sim \;$}
\def\ltsima{$\; \buildrel < \over \sim \;$}
\def\prosima{$\; \buildrel \propto \over \sim \;$}
\def\gsim{\lower.5ex\hbox{\gtsima}}
\def\lsim{\lower.5ex\hbox{\ltsima}}
\def\simgt{\lower.5ex\hbox{\gtsima}}
\def\simlt{\lower.5ex\hbox{\ltsima}}
\def\simpr{\lower.5ex\hbox{\prosima}}

\def\sigmav{\langle \sigma v \rangle}
\def\edm{\mathcal{E}_{\chi}}

\title{\boldmath Unveiling the nature of dark matter with high redshift 21 cm line experiments}

\author[a,1]{C.~Evoli,}
\author[b]{A.~Mesinger,}
\author[b]{and A.~Ferrara}

\affiliation[a]{{II.}~Institut f\"ur Theoretische Physik, Universit\"{a}t Hamburg,\\ Luruper Chaussee 149, D-22761 Hamburg, Germany}
\affiliation[b]{Scuola Normale Superiore,\\ Piazza dei Cavalieri 7, 56126 Pisa, Italy}

\emailAdd{carmelo.evoli@desy.de}
\emailAdd{andrei.mesinger@sns.it}
\emailAdd{andrea.ferrara@sns.it}

\abstract{
Observations of the redshifted 21 cm line from neutral hydrogen will open a new window on the early Universe.  By influencing the thermal and ionization history of the intergalactic medium (IGM), annihilating dark matter (DM) can leave a detectable imprint in the 21 cm signal.  Building on the publicly available 21cmFAST code, we compute the 21 cm signal for a 10 GeV WIMP DM candidate.  The most pronounced role of DM annihilations is in heating the IGM earlier and more uniformly than astrophysical sources of X-rays.  This leaves several unambiguous, qualitative signatures in the redshift evolution of the large-scale ($k\approx0.1$ Mpc$^{-1}$) 21 cm power amplitude: (i) the local maximum (peak) associated with IGM heating can be lower than the other maxima; (ii) the heating peak can occur while the IGM is in emission against the cosmic microwave background (CMB); (iii) there can be a dramatic drop in power (a global minimum) corresponding to the epoch when the IGM temperature is comparable to the CMB temperature.
These signatures are robust to astrophysical uncertainties, and will be easily detectable with second generation interferometers. We also briefly show that decaying warm dark matter has a negligible role in heating the IGM.
}

\begin{document}
\maketitle
\flushbottom

\section{Introduction}
\label{sec:intro}

The nature of dark matter (DM) is one of the crucial open questions in cosmology. DM constitutes $\sim$85 per cent of all the matter in the Universe~\cite{Komatsu:2011,PLANCK}. 
We know that DM cannot be made of ordinary matter, so particles outside the standard model (SM) of particle physics must exist (see~\cite{Bertone:2010} for a review). 
Among the many DM candidates proposed, Weakly Interacting Massive Particles (WIMPs), with particle mass $m_\chi \sim 10-1000$~GeV, are well-motivated candidates in a number of theoretical scenarios.
For this reason, although WIMPs have not been detected so far, a variety of direct and collider experiments are currently performing dedicated searches. 

Direct experiments are attempting to detect the tiny nuclear recoils induced by the interaction of DM with ordinary matter.
An annual modulation compatible with a DM signal, has been first reported by DAMA/NaI~\cite{Bernabei:1998}, hinting at the existence of a DM particle much lighter ($m_\chi \sim 5-10$~GeV) than commonly assumed for WIMPs based on supersymmetry (SUSY) neutralino scenarios. 
CoGeNT~\cite{Aalseth:2011a} and CRESST~\cite{Angloher:2012} have independently reported evidence of light-mass DM at $\sim10$~GeV.  More recently, the CDMS collaboration reported the observation by the Silicon detector of three nuclear recoil events with a best fit corresponding to a DM mass of $m_\chi = 8.6$~GeV and a WIMP-nucleon cross section consistent with CoGeNT's findings~\cite{Agnese:2013}.
On the other hand, strong constraints from the XENON100 experiment~\cite{Aprile:2012}, and more recently by LUX~\cite{LUX:2013}, appear to be incompatible with such hypotheses.
Different explanations have been invoked for these apparent discrepancies, either based on a different efficiency for the detector materials \cite[e.g.,][]{Savage:2011} or by proposing isospin-violating scenarios for the DM candidates~\cite{Feng:2011}.
Further analysis of background rates and electronics noise near the energy threshold will clarify the prospects for direct detection of light dark matter~\cite{Panci:2014}.

An alternative approach comes from collider searches, such as that at LHC, looking for the production of DM from the interactions of standard model particles. SUSY has not been confirmed by such searches. Moreover, DM candidates as light as the ones compatible with direct searches would tend to have a relic density far above WMAP-measured values~\cite{Aad:2011hh,Chatrchyan:2011zy}.

Indirect detections offer a third possibility, looking for relic signals due to DM annihilation or decay in the local Universe. The difficulty here lies in distinguishing such DM signals from astrophysical ones.
Recently, PAMELA discovered a positron excess in the galactic cosmic ray flux above 10 GeV~\cite{Adriani:2009b}.  
This was later confirmed by Fermi-LAT~\cite{Abdo:2009b} and more recently by AMS-02 on-board the International Space Station~\cite{AMS:2013}.
While this excess may be explained by $m_\chi \sim$ TeV DM annihilations or decays into electron-positron pairs~\cite{Cholis:2013}, it may also explained by astrophysical sources such as pulsars~\cite{diBernardo:2011} or middle-aged supernova remnants~\cite{Gaggero:2013}.

Alternatively, the detection of monochromatic gamma rays from DM will provide a smoking-gun signal that is very difficult to mimic by astrophysical sources~\cite{Bringmann:2012a}. The tentative discovery of such a signal at $\sim 130$~GeV in the Fermi-LAT data in a region-of-interest optimized for particular dark matter distributions towards the Galactic center has spread hopes of a long-awaited clear evidence of a signal~\cite{Bringmann2012b}. While the statistical significance of the original detection appeared to exclude a statistical fluctuation, the time evolution of the signal did not follow the expected trend~\cite{Ackermann:2013uma}. Future observations with a Cherenkov telescope (as, e.g., HESS-II) will have the required sensitivity to independently rule out or confirm the line at 130 GeV.

Indirect detections could also come from the distant Universe.  On cosmological scales, the redshifted 21 cm line from neutral hydrogen is sensitive to the density, ionization and temperature of the intergalactic medium (IGM; e.g. \cite{Furlanetto:2006r}), thus serving as powerful testbed for models of DM annihilations and decays.  Ideally, the effects of DM would be studied {\it before} the first astrophysical objects formed, during the epoch referred to as the cosmic Dark Ages ($z\gsim40$). The advantage of the Dark Ages is that they provide a clean probe of DM, without astrophysical complications.
Unfortunately, direct observations of the Dark Ages in 21 cm will likely have to wait for moon-based interferometers.

Following the Dark Ages, the first galaxies dramatically amplified the 21 cm signal through their soft UV and X-ray radiation.  This epoch is commonly referred to as the Cosmic Dawn.  The Cosmic Dawn will be easily detectable with upcoming interferometers such as the Square Kilometre Array (SKA)\footnote{\url{http://www.skatelescope.org}}~\cite{SKA:2013} and Hydrogen Epoch of Reionization Arrays (HERA)\footnote{\url{http://reionization.org}}~\cite{Pober:2014}, and possibly even with the Murchison Widefield Array (MWA)\footnote{\url{http://www.MWAtelescope.org/}}~\cite{MWA:2013} which is currently taking data~\cite{Mesinger:2014}.
 At smaller redshifts, in the reionization epoch ($z\lesssim 10$), an upper limit on the 21 cm power-spectrum was found by the  Precision Array for Probing the Epoch of Reionization (PAPER)\footnote{\url{http://eor.berkeley.edu}}: ($52$~mK)$^2$ for $k = 0.11 h$~Mpc$^{-1}$ at $z = 7.7$~\cite{Parsons:2013}, while the Low Frequency Array (LOFAR)\footnote{\url{http://www.lofar.org}} has not yet published results.

The Cosmic Dawn presents a great opportunity to search for indirect signatures of DM, before the thermal and ionization state of the IGM became governed entirely by astrophysical sources. 
 
In a previous work~\cite{Valdes:2012zv}, we found that the heating of the IGM through DM annihilations is qualitatively different from the fiducial scenario in which the IGM is heated by X-ray sources in galaxies.
DM heating suppresses the very deep absorption feature of the mean 21 cm brightness temperature, $\delta \bar{T}_{b}$, during the Cosmic Dawn.  Furthermore, DM annihilation heating is dominated by halos several orders of magnitude smaller than those hosting galaxies, whose fractional abundance evolves much slower with redshift.   
Hence DM heating is characterized by a smaller gradient $d (\delta \bar{T}_{b})/d\nu \sim 4$~mK~MHz$^{-1}$ in the range $\nu \sim 60-80$~MHz.  
Although these signatures are relatively robust, they could in principle be mimicked by ad-hoc tuning of star-formation prescriptions.

In the present work, we propose another method to discriminate between the gas heating produced by DM annihilation and that associated with astrophysical sources. The method exploits the fact that while DM heating is mainly injected from low-mass, early-forming and uniformly distributed halos, heating from stars is driven by galaxies residing in rare, biased peaks of the density field. 
While the spatial distribution of heating sources does not significantly impact the global signal, the power spectrum can be dramatically different (e.g. \cite{Sitwell:2013}).

In the following, we show how one of the most popular CDM candidates can influence the 21 cm power-spectrum. Similar results can be obtained for any DM candidate that, via annihilation or decay, has a sizable impact on IGM temperature. 

This paper is organized as follows. In \S \ref{sec:signal} we discuss how we model the cosmological 21 cm signal, including both astrophysical (\S \ref{sec:astro}) and DM (\S \ref{sec:heating}) heating sources.  We present our results in \S \ref{sec:results}, focusing on robust, qualitative signatures of DM annihilation in the evolution of the 21 cm power.  We conclude in \S \ref{sec:conclusions}.  Finally, in Appendix~\ref{sec:appendix} we estimate the heating contribution from the decay of popular Warm Dark Matter particles, finding it to be sub-dominant to astrophysical sources for reasonable scenarios.  
Unless stated otherwise, we quote all quantities in comoving units.  
We adopt a cosmology consistent with Planck results~\cite{PLANCK}:
($\Omega_\Lambda$, $\Omega_M$, $\Omega_b$, $n$, $\sigma_8$, $H_0$) $=$ ($0.68$, $0.32$, $0.049$, $0.96$, $0.83$, $67$~km~s$^{-1}$~Mpc$^{-1}$).

\section{Modeling the cosmological 21 cm signal}
\label{sec:signal}

The 21 cm line is associated with the hyperfine transition between the triplet and the singlet levels of the neutral hydrogen ground state. 
The ratio between the number densities of hydrogen atoms in the singlet ($n_0$) and triplet ($n_1$) ground hyperfine levels can be written as $n_{1}/n_{0}=3 \exp (-{T_{\star}}/T_{S})$, where $T_{\star}=0.068$~K corresponds to the transition energy, and $T_S$ is the spin temperature defined by the above expression. 

In the presence of the CMB alone, $T_S$ reaches thermal equilibrium with CMB temperature $T_\gamma = 2.73 \, (1+z)$~K on a short time-scale, making the HI undetectable either in emission or absorption. However, collisions and scattering of Ly$\alpha$ photons (the so-called Wouthuysen-Field process or Ly$\alpha$ pumping) can couple $T_S$ to the gas kinetic temperature, $T_K$, making the neutral hydrogen visible in absorption or emission depending on whether the gas is colder or hotter that the CMB \citep{Wouthuysen:1952, Field:1959, Hirata:2006}.

The spin temperature can be calculated as:
\begin{equation}
\label{eq:Ts}
{T_S}^{-1} = \frac{ {T_\gamma}^{-1} + x_{\alpha} {T_{\alpha}}^{-1} + x_c {T_K}^{-1} }{1 + x_{\alpha} + x_c}
\end{equation}
where $T_{\alpha}$ is the color temperature, which is closely coupled to $T_K$ \citep{Field:1959}, and $x_{\alpha}$ and $x_c$ are the coupling coefficients corresponding to Ly$\alpha$ scattering and collisions respectively.

The observable quantity is the differential brightness temperature between a neutral hydrogen patch and the CMB:
\begin{align}\label{Eq:dtb}
\delta T_{b} &= \frac{T_{S}-T_\gamma}{1+z}\,\tau = \nonumber\\
&\simeq\, 27 x_{\rm HI} (1+\delta_{\rm nl}) \left( \frac{H}{dv_r/dr+H} \right) \left( 1-\frac{T_\gamma}{T_S} \right) \left( \frac{1+z}{10}\frac{0.15}{\Omega_M h^2} \right)^{1/2} \left( \frac{\Omega_b h^2}{0.023} \right) \, {\rm mK}
\end{align}
where $\tau$ is the optical depth of the neutral IGM at $21(1+z)$~cm, $\delta_{\rm nl}(x, z) \equiv \rho / \bar{\rho} - 1$ is the density contrast, $H(z)$ is the Hubble parameter, $dv_r/dr$ is the comoving gradient of the LOS component of the comoving velocity, and all quantities are evaluated at redshift $z = \nu_0/\nu - 1$ (where $\nu_0$ is the 21 cm rest frame frequency).

As seen from the above, the 21 cm signal depends on: (i) the gas density $n_H$, (ii) the kinetic temperature of the gas, $T_K$; (iii) the ionized fraction $x_e$; and (iv) the Ly$\alpha$ background intensity $J_{\alpha}$ (since $x_{\alpha}\propto J_{\alpha}$; e.g. \cite{Hirata:2006}).

The ionization and thermal evolution of the IGM can be computed from:
\begin{equation} \label{eq:ion_rateacc}
\frac{dx_e}{dz} = \frac{dt}{dz} \left[ \Gamma_{\rm ion}
  - \alpha_{\rm B} C x_e^2 n_b f_{\rm H} \right] ~ ,
\end{equation}
\begin{equation} \label{eq:dTkdzacc}
\frac{dT_K}{dz} = \frac{2T_K}{1+z} + \frac{2T_K}{3 n_b}\frac{d n_b}{dz} - \frac{T_K}{1+x_e}\frac{dx_e}{dz} + \frac{2}{3 k_B (1+f_{He}+x_{e})} \frac{dt}{dz} \sum_p \epsilon_p~ ,
\end{equation}
where $n_b = \bar{n}_{b,0} (1 + z)^3 (1 + \delta_{\rm nl})$ is the total (H + He) baryon number density, $\alpha_{\rm B}$ is the recombination coefficient, $\Gamma_{\rm ion}$ is the ionization rate per baryon, $C \equiv \left<n^2\right>/\left<n\right>^2$ is the clumping factor,
$k_B$ is the Boltzmann constant, $f_{\rm H(He)}$ is the hydrogen (helium) fraction by number, and $\epsilon_p$ is the heating rate per baryon for process $p$~\cite{Chen:2004, Mesinger:2011}. 
In Eq.~\ref{eq:dTkdzacc}, the first term in the right-hand side corresponds to the Hubble expansion, the second corresponds to adiabatic heating and cooling from structure formation, and the third corresponds to the change in the total number of gas particles due to ionizations.

In absence of DM heating, the ionization rate per baryon is dominated by the astrophysical contribution, $\Gamma_{\rm ion} = \Gamma_{\rm ion,*}$, while the heating is the sum of two terms: Compton heating from CMB photons, $\epsilon_\gamma(z)$, and heating from astrophysical X-ray sources, $\epsilon_{\rm *}(z)$.

Finally, the Ly$\alpha$ background intensity receives contributions from X-ray excitation of H~I ($J_{\alpha,X}$), and direct stellar emission of photons between Ly$\alpha$ and the Lyman limit ($J_{\alpha,*}$):
\begin{equation}
\label{eq:Jalpha}
J_{\alpha,\rm tot} = J_{\alpha,X} + J_{\alpha,*}
\end{equation}
where the second term is dominant for reasonable models \cite{Mesinger:2013}.

\subsection{Astrophysical sources of ionization and heating}\label{sec:astro}

We assume that galaxies hosting UV and X-ray sources reside in atomically cooled haloes with virial temperatures $T_{\rm vir} > 10^4$ K (corresponding to halo masses of $M_{\rm halo} > 3 \times 10^7$~M$_\odot$ at $z \sim 20$). 
We assume a $10$\% efficiency of conversion of gas into stars~\citep{Evoli:2011}. Ly$\alpha$ emission is dominated by early UV sources, assumed to have a typical PopII stellar spectra~\citep{Barkana:2005}. The same galaxies produce X-ray emission following a power-law with energy index of $1.5$ with a lower limit of $300$~eV, and an X-ray efficiency corresponding to $0.2$~X-ray photons per stellar baryon. In doing this, we assume that the X-ray emission of early galaxies is similar to that observed from local star-forming ones~\cite{Mineo:2012a,Mineo:2012b,Pacucci:2014}.
The fraction of photon energy going into ionization, heating and Ly$\alpha$ emission is computed according to~\cite{Furlanetto:2010}.

There are many uncertainties associated with the astrophysical parameters mentioned above.  
A full parameter study (e.g.,~\cite{Mesinger:2014}) is beyond the scope of this work. Instead, in addition to our \emph{fiducial} model described above, we also present an \emph{extreme} model to illustrate the allowed range of uncertainties (as in \cite{Valdes:2012zv}).  In the extreme model, we take $T_{\rm vir}>10^5$ K, assume that galactic X-ray emission is more locally obscured (with only photons with energies $>900$~eV escaping the galaxy), and take an X-ray efficiency corresponding to $2000$~X-ray photons per stellar baryon. Thus in our extreme model, primordial galaxies, albeit rarer and appearing later, were much more efficient in generating hard X-rays, saturating the unresolved soft X-ray background by $z\sim 10$~\citep{Hickox:2007}.

\subsection{Dark Matter heating}
\label{sec:heating}

We now summarize how we include DM annihilations in the IGM evolution equations described in the previous section.  For further details, the reader is encouraged to see, e.g.,~\cite{Valdes:2012zv}.

High-energy electrons and positrons injected into the IGM through DM annihilations are quickly cooled down to the $\sim$keV scale mostly through inverse Compton cooling (see, e.g.,~\cite{Slatyer:2009,Evoli:2012}).
As discussed above for secondary electrons from X-ray ionizations, once the annihilation-induced shower has reached the keV energy scale it can (i) ionize the IGM; (ii) induce Lyman-excitations, and (iii) heat the plasma.

The energy deposition rate per baryon (in erg/s) for DM annihilation can be expressed as:
\begin{equation}
\label{eq:DMenergy}
\edm = \frac{1}{n_{b}} \frac{dE}{dtdV} (z) = (1+z)^{3} \frac{\Omega_{\chi}^{2}}{\Omega_{b}} \rho_{c,0} [1+B(z)] m_{p} c^{2} \frac{\sigmav}{m_{\chi}} \, ,
\end{equation}
where $m_\chi$~($m_p$) is the DM (proton) mass, and $\left< \sigma v \right>$ is the annihilation cross section averaged over the velocity distribution.
Non-linear, virialized substructures enhance the DM self-annihilation rate (with respect to the large-scale density field),  thus significantly increasing the rate of energy deposition at redshifts $z \lsim 50$.
Following~\cite{Cirelli:2009} we include non-linear structures through the parameter, $B(z)$, describing the average DM density enhancement from collapsed structures. We evaluate $B(z)$ using the Press-Schechter mass function~\cite{Press:1974} formalism and assuming a NFW~\cite{Navarro:1996} halo profile. This procedure relies on modeling the halo mass function many orders of magnitude below scales accessible through observations or cosmological simulations.  
We parameterize this uncertainty through the minimum halo mass, $M_{\rm h, min}$. 
This is usually chosen to be the free-streaming mass, $M_{\rm fs}$, which strongly depends on the assumed interaction type and mass of the DM particles. 
Modeling of the kinetic decoupling of WIMPs in the early Universe showed that the smallest halos to be formed range between $10^{-9}$ and almost $10^{-3}$ solar masses~\cite{Bringmann:2009}. In order to estimate the associated uncertainty we show results for three values of $M_{\rm h,min}$ within this range.

We find that $B(z)$ can be well-fit (at the $\sim1$\% level for $z\lsim 100$), with:
\begin{equation}\label{Eq:boost}
B(z) = \frac{b_h}{(1+z)^\delta} {\rm erfc} \left( \frac{1+z}{1+z_h} \right)
\end{equation}
with 3 free parameters, $b_h$, $\delta$ and $z_h$, whose values are given in Table~\ref{Tab:Bparams} for the different values of $M_{\rm h, min}$ adopted.
When the substructure contribution is important, the DM energy deposition is driven by $\lesssim M_\odot$ halos ~\cite{Valdes:2012zv}.  
These tiny halos can be treated as uniformly distributed on the large scales ($\gsim10$ Mpc) of interest here.

Our fiducial CDM candidate is a light WIMP with leptonic ($\mu^+\mu^-$) coupling and an annihilation cross-section compatible with thermal production $\sigmav = 10^{-26}$~cm$^3$~s$^{-1}$. Such a model has been invoked to explain a promising signal from the Galactic center and, moreover, it would be in the mass range compatible with recent claims of low-energy signals from DM direct detection if a sub-dominant hadronic component is additionally present~\cite{Hooper:2012}. 
The most recent constraints on a 10 GeV WIMP cross section, obtained by combining CMB measurements from Planck, WMAP9, ACT, and SPT, and assuming annihilation products of muons (electrons), is $\sigmav \lsim 4.3 \, (1.5) \times 10^{-26}$~cm$^3$~s$^{-1}$~\cite{Madhavacheril:2014}. 

\begin{table}[th]
\centering
\begin{tabular}{c|ccc}
\hline
\hline
M$_{\rm h, min}$ [M$_\odot$] & $b_h$ & $z_h$ & $\delta$ \\ 
\hline
$10^{-3}$ & $1.6 \times 10^5$ & $19.5$ & $1.54$ \\
$10^{-6}$ & $6.0 \times 10^5$ & $19.0$ & $1.52$ \\
$10^{-9}$ & $2.3 \times 10^6$ & $18.6$ & $1.48$ \\
\hline
\hline
\end{tabular}
\caption{Parameters used to determine the structure boost-factor (see Eq.~\ref{Eq:boost}).}
\label{Tab:Bparams}
\end{table}

Armed with eq.~\ref{eq:DMenergy}, we can express the additional contribution to the IGM evolution equations (\ref{eq:ion_rateacc}--\ref{eq:Jalpha}) from DM annihilations:
\begin{equation}\label{Eq:iondm}
\Gamma_{\rm ion,\chi} = \left[ \frac{ f_{\rm ion,HI} }{ E_{\rm 0,HI} } + \frac{ f_{\rm ion,HeI} }{ E_{\rm 0,HeI} } \right] \edm 
\end{equation}
\begin{equation}\label{Eq:heatdm}
\epsilon_\chi = f_{\rm h} \edm
\end{equation}
\noindent and
\begin{equation}\label{Eq:lyadm}
J_{\alpha,\chi} = \frac{c n_{b}(z)}{4\pi h H(z)} f_{\alpha} \edm
\end{equation}
where  $E_{\rm 0,HI}$ ($E_{\rm 0,HeI}$) is the ionization energy of hydrogen (helium), $\lambda_{\alpha}$ is the Ly$\alpha$ wavelength, and $f_{\rm ion,HI}$, $f_{\rm ion, HeI}$, $f_{\rm h}$, $f_{\alpha}$ are the energy fractions deposited into hydrogen ionization, helium ionization, heating and Ly$\alpha$ excitations, respectively.  These energy fractions are computed according to \cite{Evoli:2012}, whose results are consistent within 20\% to the analytic calculation taking into account photon redshifting from \cite{Slatyer:2009}.

\subsection{Simulations}\label{sec:code}
Since we are simulating an interferometric signal, we must model inhomogeneous ionizations and heating, integrating the evolution of cosmic structures and radiation fields along the light cone.  For this purpose we use the public code 21cmFAST\footnote{http://homepage.sns.it/mesinger/Sim.html}, which generates cosmological density, and astrophysical radiation fields.  Our simulation boxes are $600$~Mpc on a side, with a resolution of $400^3$. We modify the code to include the additional (homogeneous) contribution from DM annihilations, as described above. For further details and tests of the code, interested readers are encouraged to see~\cite{Mesinger:2007,Zahn:2011,Mesinger:2011}.

\section{Results}
\label{sec:results}

\begin{figure}
\centering
\includegraphics[width=0.49\textwidth]{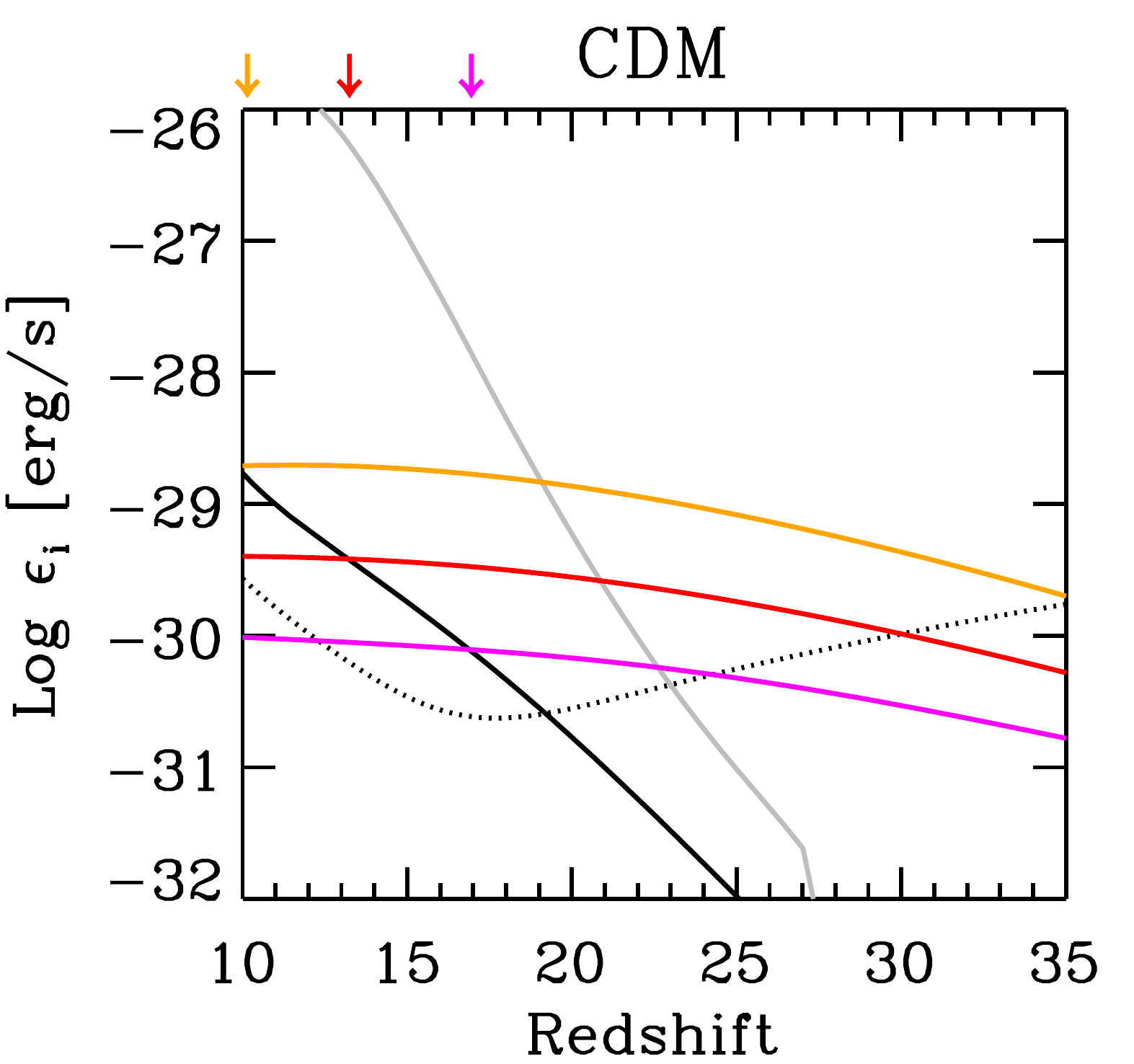}
\hspace{\stretch{1}}
\includegraphics[width=0.49\textwidth]{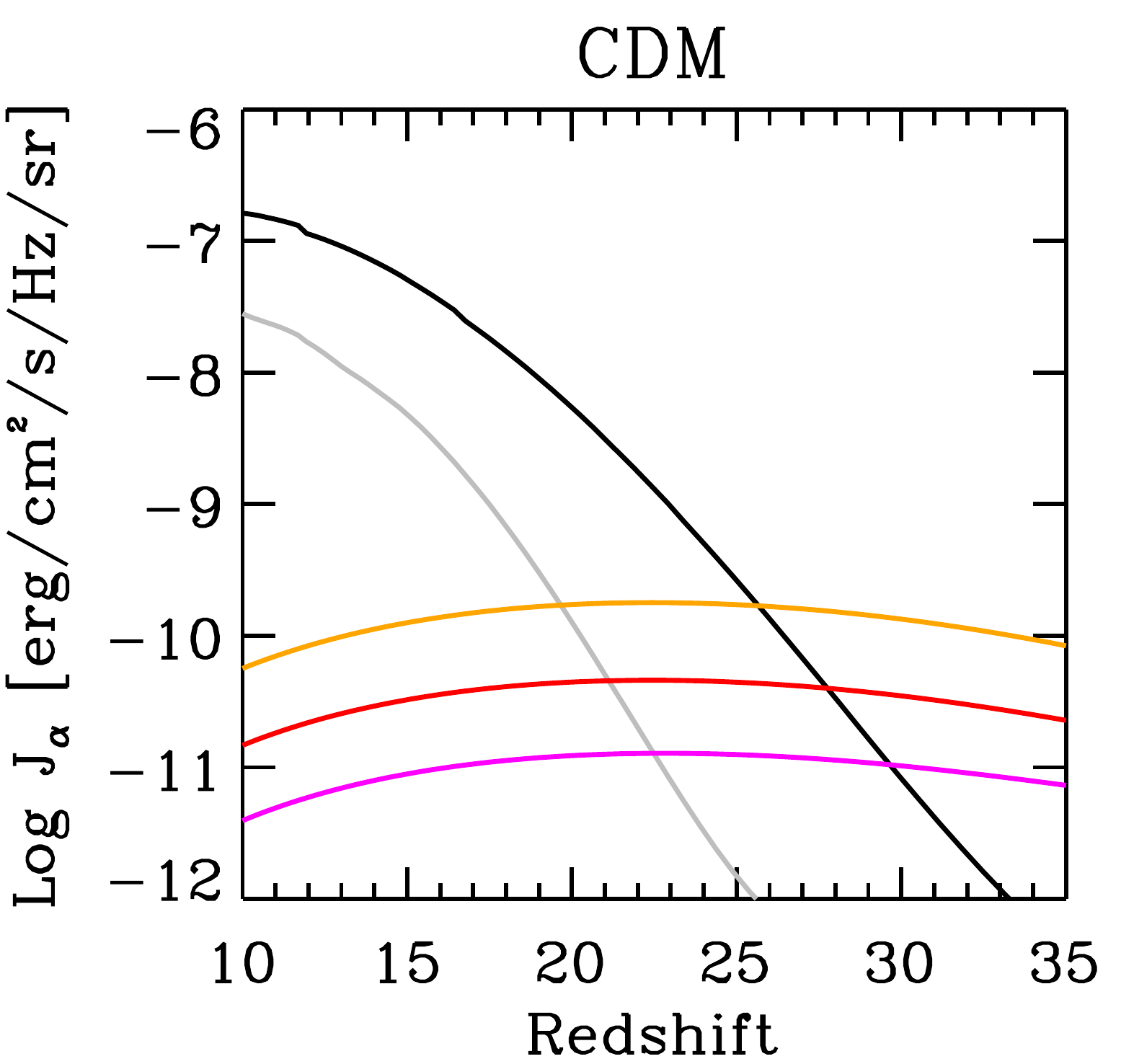}
\caption{\small Heating rates (left) and intensity of the Ly-$\alpha$ radiation background (right) produced by CDM annihilations assuming various values of $M_{\rm h, min}$.  For comparison, we also show values corresponding to astrophysical sources in the fiducial (black-solid curve) and extreme (gray-solid curve) model, as well as the adiabatic cooling rate (black-dotted curve).  
The arrows on the top side of the plot indicate the transition between DM and astrophysical sources as dominant heating source (assuming fiducial astrophysics).}\label{fig:heating}
\end{figure}

\subsection{Rates}

Before presenting estimates of the 21 cm signal, we compare the relative contribution of astrophysical sources and DM annihilations to the evolution of the IGM.  In the right panel of Figure~\ref{fig:heating} we show the heating rates per baryon from DM annihilations, together with astrophysical X-ray heating.
As already noticed in~\cite{Valdes:2012zv}, the heating rate from astrophysical sources of X-rays evolves much steeper with redshift than the DM heating. This is expected since the fractional increase of the collapsed fraction in $\lesssim 1$~M$_\odot$ haloes, which drive the DM heating, is much slower than the fractional increase in the high-end tail of the mass function (i.e.~the haloes which host the first galaxies). 
This translates to a shallower redshift gradient of the mean signal, for models in which DM annihilations dominate the IGM evolution~\cite{Valdes:2012zv}.

For our three fiducial choices of $M_{\rm h, min}$, the heating from fiducial astrophysical sources exceeds that of DM annihilations at $z<10$ ($M_{\rm h, min} = 10^{-9}$), $z<13$ ($M_{\rm h, min} = 10^{-6}$) and $z<17$ ($M_{\rm h, min} = 10^{-3}$).
 In the extreme model, the astrophysical X-ray heating is dominant already at earlier redshifts ($z\sim 20$).

From the right panel of Figure~\ref{fig:heating}, we see that the fiducial Ly$\alpha$ background from astrophysical sources dominates over the DM-induced one up to $z \lsim$ 25--30 (note that the extreme astrophysical model has a lower soft UV emissivity than the fiducial one). Hence, compared to astrophysical sources, DM annihilations have a larger contribution to heating the IGM than they do to WF coupling.  As we see below, this means that DM annihilations are not coupling the 21 cm spin temperature with the gas temperature efficiently, except for the lowest $M_{\rm h,min}$ model.

\begin{figure}
\centering
\includegraphics[width=0.6\textwidth]{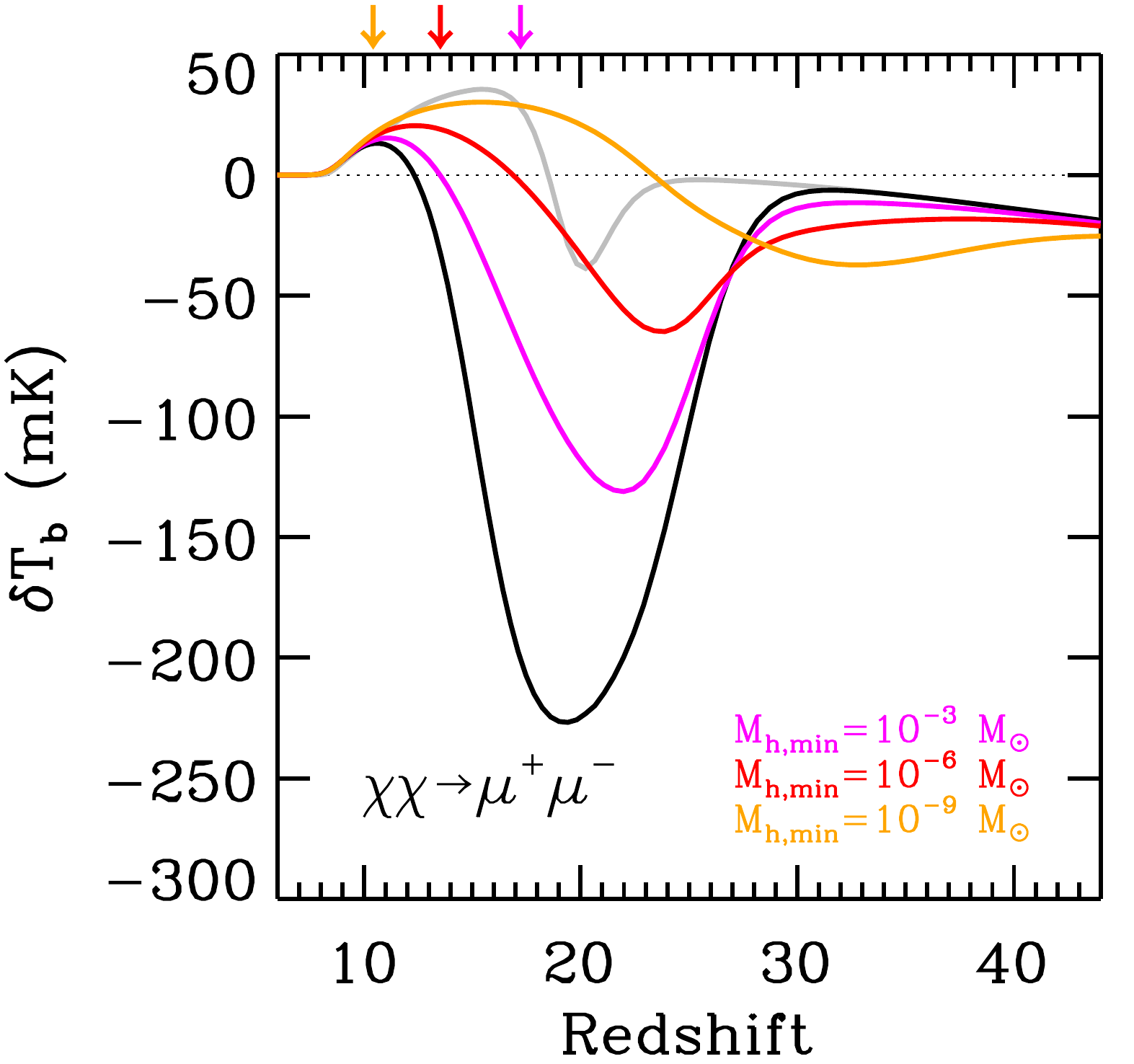}
\caption{\small Redshift evolution of the average 21 cm brightness temperature offset from the CMB.
The arrows on the top side of the plot indicate the transition between DM and astrophysical sources as the dominant heating source.}\label{fig:Tb}
\end{figure}

\subsection{Global signal}

Before moving on to the power spectrum, in Figure~\ref{fig:Tb} we show the evolution of the mean 21 cm brightness temperature, highlighting the relevant epochs.
Without DM annihilations (black-solid line), the global 21 cm signal has the following milestones (e.g.,~\cite{Furlanetto:2006r}):
\begin{enumerate}
\item At much higher redshifts ($z \gsim 50$) than shown in Figure~\ref{fig:Tb}, the IGM is dense enough that the spin temperature is uniformly coupled to the gas kinetic temperature because of collisions. Therefore $T_K = T_S \lsim T_\gamma$. The IGM cools adiabatically faster than the CMB, making $\delta T_b$ negative.
\item As the IGM becomes less dense, the spin temperature starts to decouple from the kinetic temperature, and begins to approach the CMB temperature again, $T_K < T_S \lsim T_\gamma$. As a consequence, $\delta T_b$ starts rising towards zero. This epoch corresponds to $z \gsim 30$ in Figure~\ref{fig:Tb}.
\item As soon as the first astrophysical sources turn on, their soft UV photons quickly establish a Ly$\alpha$ background, which again couples $T_S$ to $T_K$, through the WF (or Ly$\alpha$ pumping) mechanism.  Again, $\delta T_b$ is negative, potentially reaching its largest absolute value. In Figure~\ref{fig:Tb} the corresponding absorption feature reaches $\delta T_b \lsim -200$ at around $z \sim 20$.
\item Subsequently, X-rays from galaxies heat the IGM. As the gas temperature surpasses $T_\gamma$, the 21 cm signal changes from absorption to emission (at around $z \sim 12$ here).
\item Finally, the IGM is reionized, and the signal again approaches zero.
\end{enumerate}
    
This scenario would change considerably if we allow for DM heating.
Firstly, the additional heating at very high redshifts can place $T_K$ on a higher adiabat already during the collisional decoupling era.   
This dampens the depth of the absorption trough at $z\sim20$--30.  
Secondly, if DM annihilations contribute to Ly$\alpha$ pumping and heating, the signal would show a more gradual evolution during the absorption epoch.

The main result is the damping and smoothing of the absorption feature observed at $16<z<30$ in the fiducial model. 
In particular, depending on the assumed $M_{\rm h, min}$, the variation in $\delta T_b$ at the minimum is predicted to be of $110$, $190$ and $200$ mK for $M_{\rm h, min} = 10^{-3}$, $10^{-6}$ and $10^{-9}$~$M_\odot$ respectively. 

However, a similar qualitative trend is also present in the \emph{extreme} astrophysical model described in Sec.~\ref{sec:astro}, in which we allow for an enhanced production of hard X-rays.  
This partial degeneracy makes is difficult to extract a robust signature of DM annihilation heating from the global signal.
In the next Section we show that the different spatial distribution of the relevant heating sources allows to discriminate between the two scenarios.

\subsection{Power-spectrum}

As our main observable, we use the spherically averaged power spectrum:
\begin{equation}
P_{21} \equiv \frac{k^3}{2\pi^2V} \delta \bar{T}_b(z)^2 \langle |\delta_{21}({\bf k},z)|^2\rangle_k
\end{equation}
where $\delta_{21}({\bf x},z) \equiv \delta T_b({\bf x},z)/\delta \bar{T}_b-1$.  Our default power spectrum bin width is $d\ln k = 0.5$.

\begin{figure*}
\centering
\includegraphics[width=0.6\textwidth]{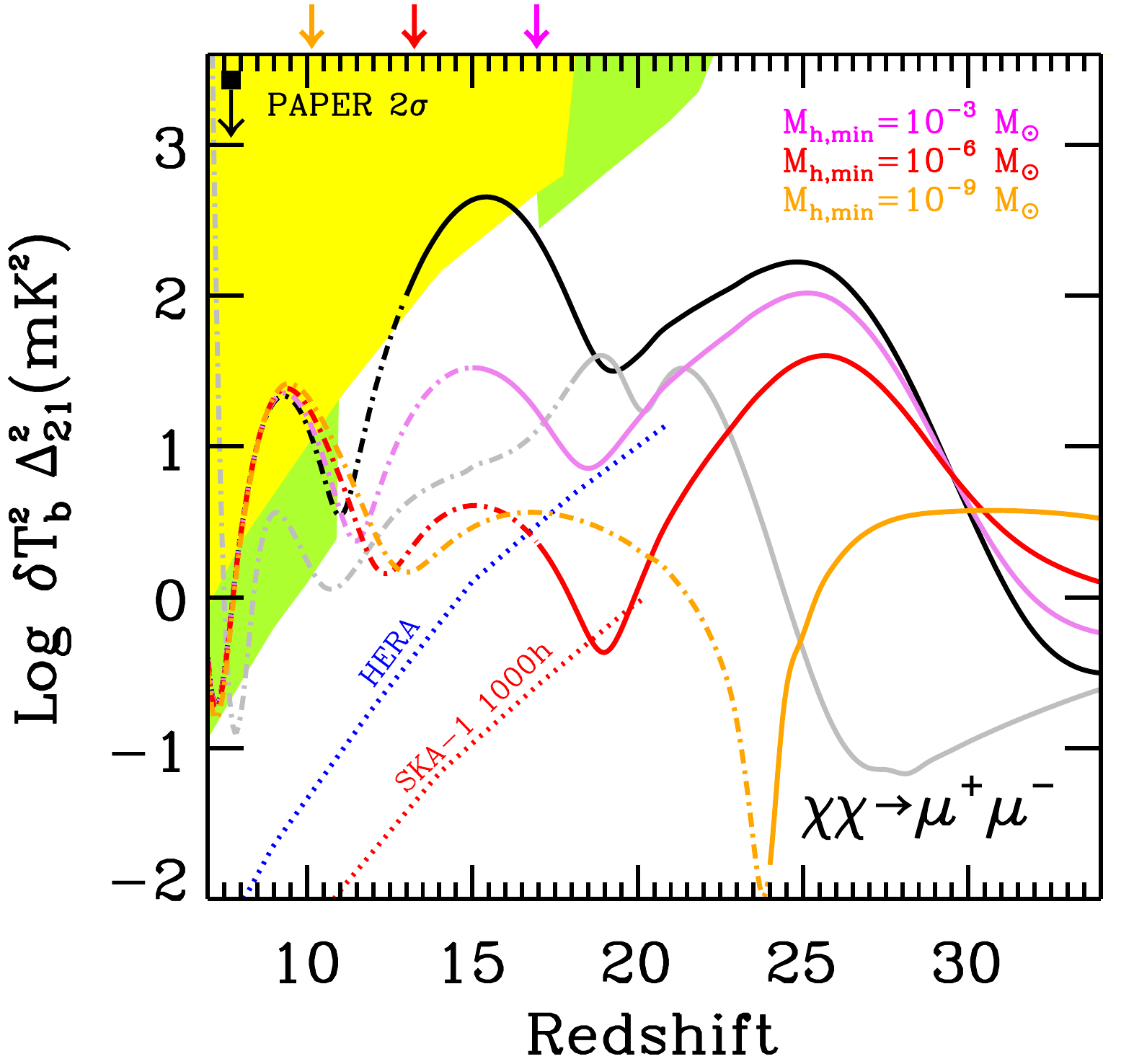}
\caption{\small Evolution of the 21 cm power at $k = 0.1$~Mpc$^{-1}$ for all of the considered DM and astrophysical X-ray models (solid line if the corresponding mean signal is in absorption, dashed-dotted line if in emission). The shaded areas correspond to the sensitivity regions calculated in~\cite{Mesinger:2013b} for the experiments: MWA-128T (yellow), LOFAR (green). SKA (HERA) single beam 1000h sensitivity limit is plotted as a dotted red (blue) line.
The arrows on the top side of the plot indicate the transition between DM and astrophysical sources as dominant heating source.}\label{fig:ps}
\end{figure*}

In Figure~\ref{fig:ps}, we show the redshift evolution of the $k=0.1$~Mpc$^{-1}$ mode of the 21 cm power spectra for the same models shown in Figure~\ref{fig:Tb}.  This scale roughly corresponds to the narrow window of $k$-space accessible to the first generation interferometers (e.g. \cite{Chapman:2012,Pober:2013}).
In order to predict the detectability of the signal, we also show 1$\sigma$ thermal noise corresponding to a 1000h, single-beam, observation with some upcoming and current instruments (taken from~\cite{Mesinger:2013b}).

The redshift evolution of the large-scale 21 cm power is characterized by three peaks, corresponding to (from high to low $z$): (i) WF coupling (fluctuations in $J_{\alpha}$); (ii) X-ray heating (fluctuations in $T_K$); (iii) reionization (fluctuations in $x_i$).
The earlier peaks (especially the X-ray heating one) are larger, sourced by the larger available contrast during the absorption epochs.
Increasing the X-ray efficiency mainly shifts the X-ray heating peak to earlier epochs. If X-ray heating occurs early enough, it overlaps with the Ly$\alpha$ pumping epoch, decreasing the associated peak in power \cite{Mesinger:2013b}.  Similarly, if galactic sources are characterized by hard ($\gsim1$keV) X-rays with long mean free paths, the heating would be much more uniform, again decreasing the associated peak in power \cite{Mesinger:2013, Pacucci:2014}. 

DM annihilations can impact this picture in several ways.  The annihilation products from DM have a larger fractional contribution (compared with astrophysical sources) to the IGM heating rate than the Ly$\alpha$ coupling (see Fig. \ref{fig:heating}).
Hence, the DM imprint is more evident in the middle peak corresponding to the temperature fluctuations during the heating era.

If DM annihilations were indeed important contributors, IGM heating would (i) start early and (ii) be more uniform, as discussed above.  
These two attributes of DM heating have {\it three, robust, qualitative signatures in the 21 cm power spectrum evolution}. 
Firstly, the early heating raises the IGM temperature to values in excess of those reached through adiabatic cooling.  
Therefore the contrast between the higher $T_K$ and $T_\gamma$ is reduced.  
Moreover, the uniformity of the DM heating does not source large-scale temperature fluctuations.
The resulting temperature contrast during the heating epoch and the associated 21 cm peak in large-scale power amplitude is greatly reduced.   
In fact, with DM annihilations the heating (middle) peak in the evolution of 21 cm power can be {\it lower than the other two peaks} (associated with WF coupling and reionization).  
This qualitative signature is very difficult to mimic with astrophysics, requiring unrealistically hard X-ray spectra (e.g. \cite{Pacucci:2014}).

\begin{figure}
\centering
\includegraphics[width=0.58\textwidth]{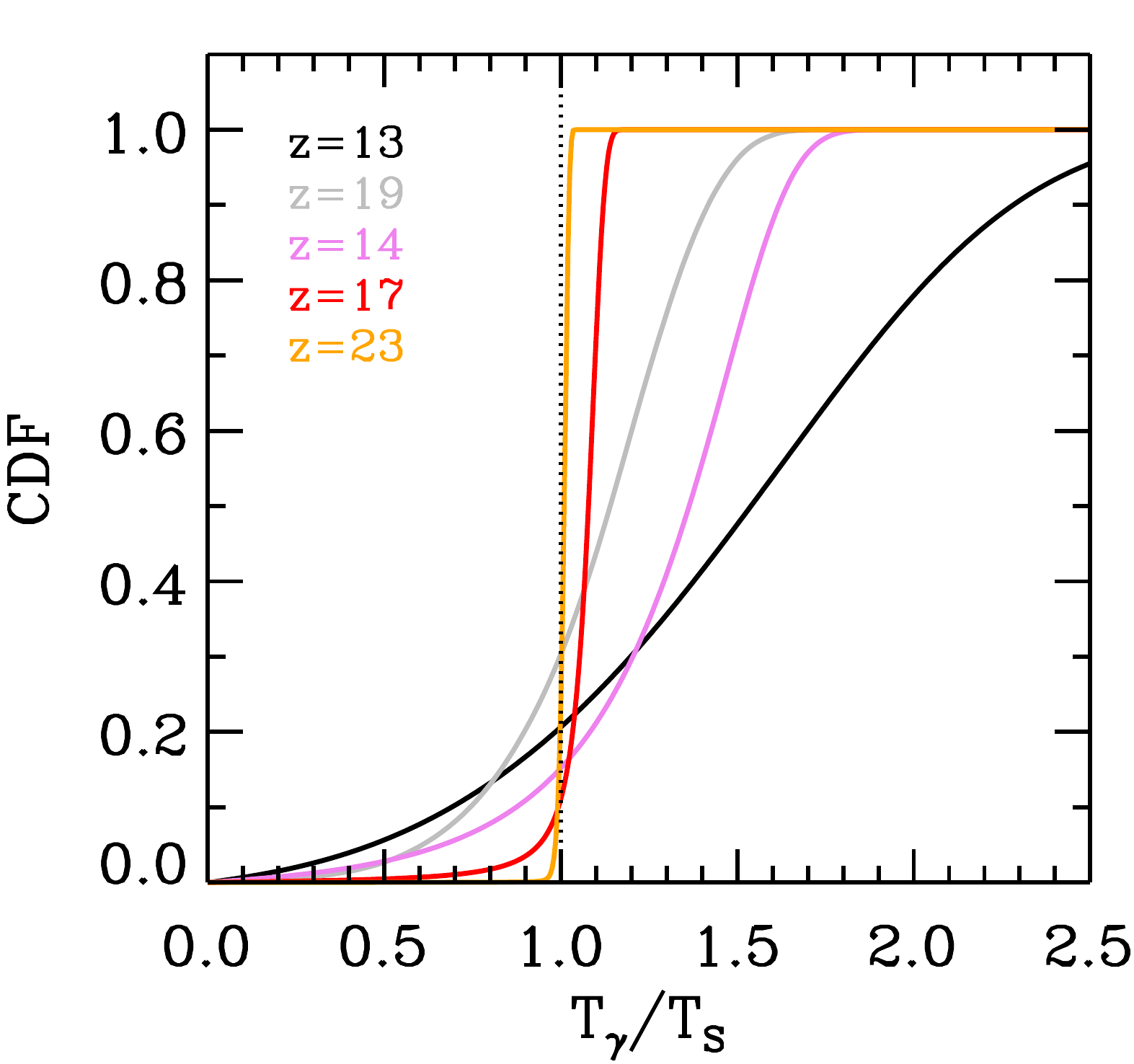}
\caption{\small Fraction of the IGM having $T_\gamma/T_S$ less than the given value. For each model, the CDFs correspond to the redshift at which $T_S \sim T_\gamma$ (see legend). Regions with $T_\gamma/T_S < 1$ are visible in emission, while the others are visible in absorption against the CMB.
}\label{fig:THist}
\end{figure}

Secondly, models in which DM annihilations heat the IGM to temperatures $T_S>T_\gamma$ {\it alone} (i.e. without astrophysical sources), result in a dramatic {\it drop in large-scale power between the WF and heating epochs.}  This is evident in our model with $M_{\rm h,min} = 10^{-9}$~M$_\odot$.  In this model, the IGM is pre-heated by annihilations before the astrophysical sources start to contribute.  Because this heating is uniform, there are no associated large-scale temperature fluctuations to dominate the 21 cm power when $T_S=T_\gamma$ and the mean signal is $\delta T_b=0$ (e.g.,~\cite{Pritchard:2006}).  This is further illustrated in Fig. \ref{fig:THist}, where we plot the cumulative distribution functions (CDFs) of $T_\gamma/T_S$, at the redshift when $T_S \sim T_\gamma$ in each model.  Models in which the DM annihilations have a stronger contribution to heating result in smaller spatial fluctuations in the IGM temperature.

The third imprint of DM annihilations on the 21 cm signal is even more unambiguous.
Even if uniform heating by DM dominates, patchy heating driven by astrophysical sources will eventually imprint (some) large-scale temperature fluctuations.  The middle peak in the 21 cm power evolution is always set by astrophysical sources (note that in Fig.~\ref{fig:ps} it always occurs roughly at the same redshift for a given astrophysical model).
However, since DM annihilations pre-heat the IGM, the X-ray heating peak occurs {\it when the IGM is already in emission against the CMB.}  This cannot be mimicked by astrophysics!  Without DM heating, the peak in 21 cm power amplitude associated with X-ray heating always occurs when the IGM is {\it in absorption} against the CMB, when the temperature distribution is broadest.
This signature represents a further motivation for combined all-sky and interferometric observations of the 21 cm signal.

\section{Conclusions}
\label{sec:conclusions}

We investigate the impact of DM annihilations on the cosmological HI 21~cm signal.
As our fiducial DM candidate we take a 10~GeV WIMP annihilating in $\mu^+\mu^-$ with a thermal cross-section.
In this context annihilation in leptons provides the largest signal with respect to other (e.g., hadronic) annihilation channels.

Building on the public 21cmFAST code, we compute the redshift evolution of the large-scale ($k\approx0.1$ Mpc$^{-1}$) 21cm power during the Cosmic Dawn.
We find that it is easiest to isolate the imprint of DM annihilations during the epoch when the IGM was heated to temperatures exceeding that of the CMB.  The DM annihilation contribution to this heating epoch can easily exceed that of astrophysical X-ray sources.

Astrophysical sources are hosted by highly-biased galaxies, whose clustering imprints large-scale temperature fluctuations.  On the other hand,
DM annihilations occur relatively uniformly, driven by small-mass, early appearing halos. This means that they uniformly heat the IGM at very high redshifts ($z\gsim20$), providing a temperature floor that damps the subsequent galaxy-driven fluctuations.
This results in two robust signatures of DM heating in the redshift evolution of the amplitude of the large-scale 21 cm power: \\
\begin{itemize} 
\item the second local maximum (associated with IGM heating) is lower than the other two (corresponding to reionization and Ly$\alpha$ pumping);
\item this local maximum occurs when the 21 cm global signal is in emission with respect to the CMB.
\end{itemize} 
While the former of these could be mimicked by astrophysical heating with very hard X-rays (a highly unlikely scenario; \cite{Pacucci:2014}), the later {\it cannot be mimicked by astrophysics}.  These signatures can easily be detected by second-generation instruments, such as HERA and SKA.

Moreover, models in which DM annihilations dominate IGM heating (e.g. our $M_{\rm h, min} =10^{-9}$~$M_\odot$ model), are also characterized by a dramatic drop in the power spectrum amplitude at the redshifts when $T_S \sim T_\gamma$.  In such scenarios, the DM annihilations heat the IGM well before the clustering of galactic X-ray sources manages to imprint large-scale temperature fluctuations in the IGM.  A null detection in that redshift range by interferometers it will provide exciting evidence of DM annihilation driven pre-heating.

More generally, the same qualitative trends are expected for any exotic model providing a uniform heating with a $\sim 10^{-30}$~erg/s heating rate at epochs before the first galaxies. This possibility is within a number of the DM supersymmetric candidates proposed in natural extensions of the SM. Furthermore, non-thermal DM production or some other mechanism, such as Sommerfeld enhancement~\cite{Slatyer:2009}, can motivate considerably higher annihilation cross-sections than the thermal relic value considered here.

\acknowledgments

C.~E.~acknowledges a visiting grant from SNS where part of this work has been carried out,
and support from the ``Helmholtz Alliance for Astroparticle Physics HAP'' funded by the Initiative and Networking Fund of the Helmholtz Association. 

\appendix

\section{Heating by WDM decay}\label{sec:appendix}

Over the last decade, warm dark matter (WDM) has been repeatedly proposed as an alternative to CDM. For a given value of $m_{\chi}$ and associated degree of freedom $g_{d}$, DM particles can free-stream over the scale (assuming a relativistic thermal relic)
\begin{equation}\label{Eq:threlic}
r_{\rm fs} = 57.2 \, {\rm kpc} \, \left( \frac{\rm keV}{m_\chi} \right) \left( \frac{100}{g_d} \right)^{1/3} \, \, ,
\end{equation}
thereby suppressing structures on smaller scales, and helping explain observations of local dwarf galaxies.

A popular WDM candidate is the $\sim$keV sterile neutrino~\cite{Dodelson:1994}.
Such a particle is predicted to decay into a lighter neutrino and a photon producing a narrow line in the X-ray spectrum at an energy equal to half the mass of the decaying neutrino.
The released $\sim$keV photon quickly deposits its entire energy into the IGM~\cite{Valdes:2008}, at a rate of: 
\begin{equation}
\edm = \frac{1}{n_{b}} \frac{dE}{dtdV} (z) = \frac{\Omega_\chi}{\Omega_b} \frac{m_p c^2}{\tau_\chi} \,
\end{equation}
where $\tau_\chi$ is the particle life-time and it is dependent on the mass and on the contribution to relic abundance, $\Omega_s$, as given by numerical simulations~\cite{Abazajian:2001}:
\begin{equation}\label{eq:decay}
\tau_\chi^{-1} = \Gamma (\nu_{s} \rightarrow \gamma \nu_{\alpha}) \sim 5.9 \times 10^{-30} \left( \frac{\Omega_{\nu_{s}}h^2}{0.12} \right) \left( \frac{m_\chi}{\rm keV}\right)^{3.2} \, {\rm s}^{-1}  
\end{equation}
We  assume that the sterile neutrino density dominates the observed DM ($\Omega_\chi$) abundance, hence $\Omega_{\nu_s} = \Omega_\chi$, and that it is a thermal relic.
Current measurements place limits of $m_{\rm wdm} \gsim 1-3$~keV~\cite{Barkana:2001,deSouza:2013,Kang:2013,Pacucci:2013,Viel:2013}, with various degrees of astrophysical degeneracy.

\begin{figure}
\centering
\includegraphics[width=0.6\textwidth]{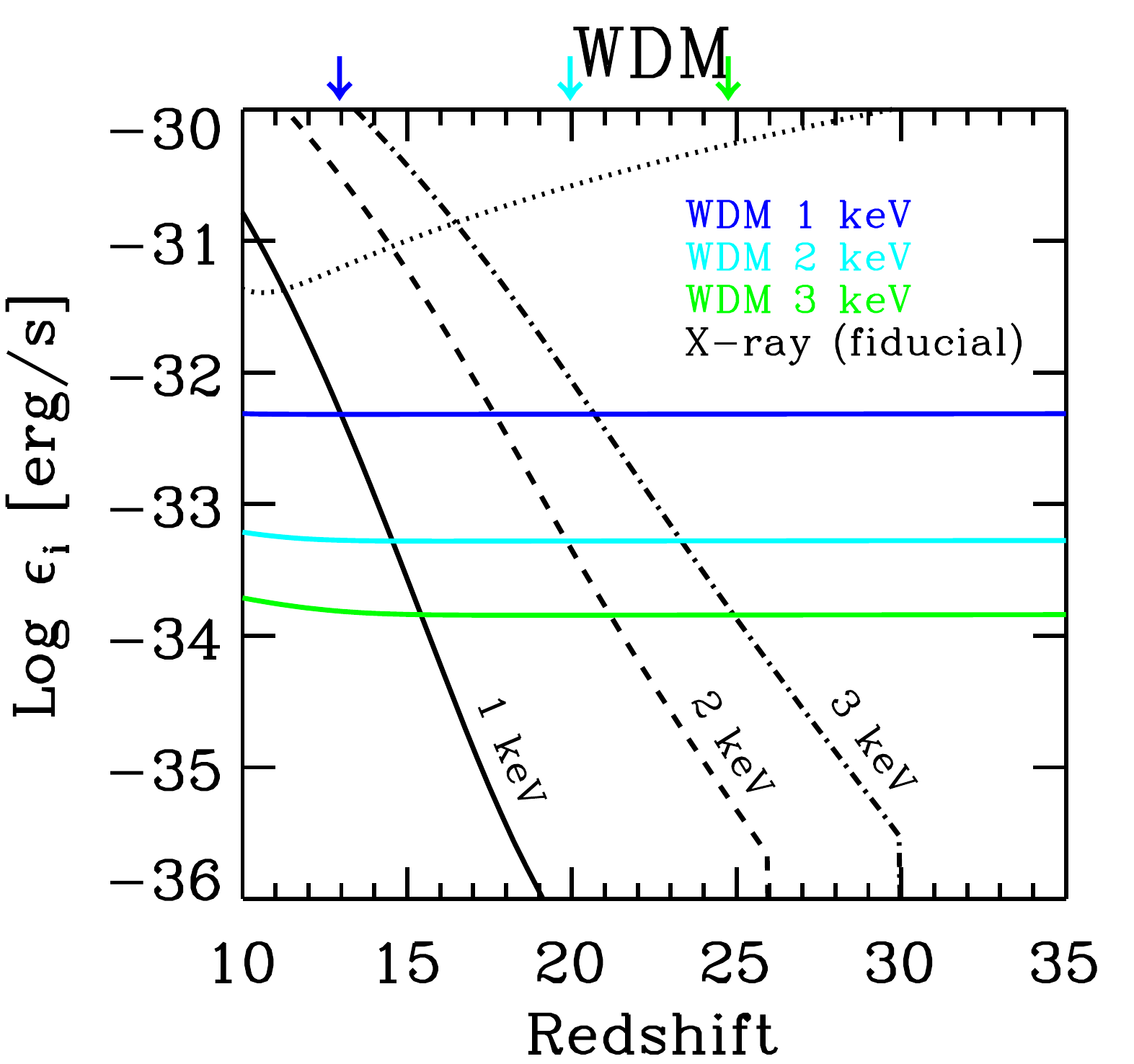}
\caption{\small Heating rates produced by WDM decay and compared with astrophysical heating rates in the fiducial astrophysics model.  The adiabatic cooling rate is shown with a dotted curve. Note that also the astrophysical heating rates depend on the particle mass since the abundance of X-ray halos is affected by the power-spectrum suppression induced by WDM free-streaming. Heating through WDM decays is always sub dominant to adiabatic cooling.}
\label{fig:heating_WDM}
\end{figure}

We show the associated heating rates in Fig.~\ref{fig:heating_WDM}.  Unlike for CDM annihilations, the WDM decay heating rate is always lower than the adiabatic cooling rate.  
Hence WDM decays do not have an notable imprint on the 21 cm signal.

Note that in addition to energy injection into the IGM through decays, WDM modifies the 21 cm signal through the associated suppression of halos hosting early galaxies.  This imprint could be detectable with upcoming interferometers~\cite{Sitwell:2013}.

\bibliographystyle{JHEP}
\bibliography{medea}

\end{document}